# Growth Window of Ferroelectric Epitaxial Hf$_{0.5}$Zr$_{0.5}$O$_2$ Thin Films

*Jike Lyu, Ignasi Fina, Raul Solanas, Josep Fontcuberta, and Florencio Sánchez\**

Institut de Ciència de Materials de Barcelona (ICMAB-CSIC), Campus UAB, Bellaterra 08193, Barcelona, Spain

*Email: fsanchez@icmab.es

ABSTRACT The metastable orthorhombic phase of hafnia is generally obtained in polycrystalline films, whereas in epitaxial films its formation has been much less investigated. We have grown Hf$_{0.5}$Zr$_{0.5}$O$_2$ films by pulsed laser deposition and the growth window (temperature and oxygen pressure during deposition, and film thickness) for epitaxial stabilization of the ferroelectric phase is mapped. The remnant ferroelectric polarization, up to ~24 μC/cm$^2$, depends on the amount of orthorhombic phase and interplanar spacing and increases with temperature and pressure for a fixed film thickness. The leakage current decreases with an increase in thickness or temperature, or when decreasing oxygen pressure. The coercive electric field (E$_C$) depends on thickness (t) according the E$_C$ - t$^{-2/3}$ scaling, which is observed by the first time in ferroelectric hafnia, and the scaling extends to thickness down to around 5 nm. The proven ability to tailor functional properties of high quality epitaxial ferroelectric Hf$_{0.5}$Zr$_{0.5}$O$_2$ films paves the way toward understanding their ferroelectric properties and prototyping devices.

1. INTRODUCTION

Doped hafnium oxide, with robust ferroelectricity at room temperature and fully compatibility with CMOS fabrication processes, is expected to have big impact in microelectronics.[1-3] The polar orthorhombic metastable phase of hafnium oxide that appears in properly doped thin films shows ferroelectricity.[4] When hafnium oxide is doped with Zr, the ferroelectric polarization is high in a broad composition range around $Hf_{0.5}Zr_{0.5}O_2$ (HZO).[3,5] The orthorhombic phase is commonly obtained by annealing an amorphous doped hafnia film inserted between top and bottom TiN electrodes. The annealing makes the film polycrystalline, and the metastable orthorhombic phase coexists with paraelectric phases. The relative amount of the orthorhombic phase and the ferroelectric properties depend on the annealing conditions and film thickness.[3,6-13]

The ferroelectric orthorhombic phase can be also stabilized in epitaxial films.[14-21] In epitaxial films, the orthorhombic phase is generally formed during deposition at high temperature, without need of annealing.[14-20] Epitaxial films are of high interest for better understanding of the properties of ferroelectric hafnia, as well as for prototyping devices with ultrathin films or having small lateral size, for which the higher homogeneity of epitaxial films respect to polycrystalline films is an advantage. In spite of the evident interest, epitaxial ferroelectric hafnia is still in a nascent state, and few groups have reported epitaxial films on YSZ,[14-16,21,22] oxide perovskite,[17,19,23] and Si [18,20] substrates. Up to now, the epitaxial films have been grown by pulsed laser deposition (PLD), and only the influence of thickness has been discussed.[17,21] The effect of deposition parameters on structural and ferroelectric properties, which is of pivotal importance for further development of epitaxial films of ferroelectric hafnia, is unreported. Here, we present a detailed study of epitaxial growth of HZO on $SrTiO_3$ (STO) substrates. Three series of samples were prepared varying deposition temperature, oxygen pressure and thickness. The growth window of epitaxial ferroelectric hafnia films is mapped, permitting the control of the structural and functional properties by selection of deposition parameters and film thickness. We find that growth parameters and thickness determine the relative amount of coexisting phases in the film, and the lattice strain of orthorhombic HZO phase in a range wider than 3%, having the films extremely flat surface. The electrical properties can be tailored, with low leakage around $10^{-7}$ A/cm$^2$ (at 1 MV/cm) in the most insulating films, and the remnant polarization ranging from negligible value up to around 24 μC/cm$^2$. The coercive field – thickness$^{-2/3}$ scaling, often observed in ferroelectric perovskites, is reported by the first time for ferroelectric hafnia films.

2. EXPERIMENTAL

Bilayers combining ferroelectric HZO film on $La_{2/3}Sr_{1/3}MnO_3$ (LSMO) bottom electrode were grown on STO(001) in a single process by PLD (248 nm wavelength). The LSMO electrodes, 25 nm thick, were deposited at 5 Hz repetition rate, substrate temperature $T_s$ = 700 °C (measured by a thermocouple inserted in the middle of the heater block), and dynamic oxygen pressure $PO_2$ = 0.1 mbar. Three series of samples were prepared varying deposition conditions of HZO (see a schematic in Figure S1): $T_s$-series, $PO_2$-series, and a thickness series. In $T_s$-series, HZO was deposited varying $T_s$ from 650 to 825 °C, under fixed conditions of $PO_2$ = 0.1 mbar and number of laser pulses (800 p, HZO thickness t = 9.2 nm). HZO was deposited in $PO_2$-series varying $PO_2$ in the 0.01-0.2 mbar range, at fixed $T_s$ = 800 °C and 800 laser pulses. The thickness of films in $PO_2$-series was in the 8-11 nm range (Figure S2). In t-series HZO films of varied thickness were prepared at $T_s$ = 800 °C and $PO_2$ = 0.1 mbar, controlling the thickness (in the 2.3 – 37 nm range) with the number of laser pulses (from 200 to 3600). At the end of the deposition, samples were cooled under 0.2 mbar oxygen pressure. Structural characterization was performed by X-ray diffraction (XRD) using Cu Kα radiation and atomic force microscopy (AFM) in dynamic mode. Platinum top electrodes, 20 nm thick and 20 µm in diameter, were deposited by dc magnetron sputtering through stencil masks. Ferroelectric polarization loops at frequency of 1kHz and current leakage were measured in top-bottom configuration (grounding the bottom electrode and biasing the top one)[24] at room temperature using an AixACCT TFAnalyser2000 platform. Leakage contribution to the polarization loops was minimized using dynamic leakage current compensation (DLCC) standard procedure.[25,26] The presence of a large dielectric contribution is manifested by the substantial slope of the polarization loops, which is common to HZO films.[19,20,27,28]

3. RESULTS AND DISCUSSION

We first address the effect of the deposition temperature ($T_s$-series) on the crystallinity of the HZO films. The XRD θ-2θ scans (Figure 1a) show (00l) reflections of STO and LSMO, and diffraction peaks in the 2θ range of 27 – 35° corresponding to HZO. The highest intensity HZO peak is the (111) reflection of orthorhombic HZO (o-HZO) at around 30°. Reflections of the monoclinic (m) phase, (-111) at 2θ around 28.5° and (002) at 2θ around 35°, usual in polycrystalline films,[27] are not detected. Laue fringes (some of them marked with vertical arrows) can be observed around o-HZO(111). Simulation of the interference fringes is presented in Figure S3. The intensity of the o-HZO(111) peak, normalized to that of the LSMO(002) peak, increases monotonously with $T_s$ (Figure 1b). Further XRD characterization was performed using a two-dimensional (2D) detector. The 2θ-χ frame around χ = 0° of the $T_s$ = 800 °C film is shown in Figure 1c. The monoclinic HZO(002) reflection is present, with broad intensity distribution along χ that indicates high mosaicity. The o-HZO(111) reflection is bright in spite of the low film thickness (t = 9 nm), and the narrow spot around χ = 0° is a signature of epitaxial ordering. The ϕ-scan around asymmetrical o-HZO(-111) reflections (Figure 1d) confirms that the o-HZO phase on LSMO/STO(001) is epitaxial, and the four sets of three o-HZO(-111) peaks indicate that it presents four crystal domains. The same intriguing epitaxial relationship and domain structure was observed in thicker o-HZO films.[19] The substrate temperature has an impact on the out-of-plane lattice parameter of o-HZO. The vertical dashed line in Figure 1a marks the position of the o-HZO(111) peak in the $T_s$ = 825 °C film. The peak shifts moderately to higher angles with substrate temperature. The dependence of $d_{o-HZO(111)}$ on $T_s$ (Figure 1e) shows that lattice spacing $d_{o-HZO(111)}$ decreases from 2.979 Å ($T_s$ = 650 °C) to 2.959 Å ($T_s$ = 825 °C), which corresponds to a contraction of 0.67%.

All the $T_s$-series films have very flat surfaces. Topographic AFM images of the $T_s$ = 650 and 825 °C films are shown in Figure 2a and 2b, respectively, and the corresponding images of all the films in the $T_s$ series are in Figure S4-1. The $T_s$ = 650 °C film is particularly flat, with root mean square (rms) roughness of 0.21 nm. The $T_s$ = 825 °C film is slightly rougher, but the rms roughness being as low as 0.36 nm. There are terraces and steps[29] in some of the films (Figure S4-1). The dependence of the rms roughness on $T_s$, with rms in the 0.21 - 0.36 nm range, is shown in Figure 2c.

Figure 3 summarizes the influence of deposition oxygen pressure ($PO_2$-series) on the crystallinity of the films. There are not HZO diffraction peaks in the $PO_2$ = 0.01 mbar film (Figure 3a), whereas in the $PO_2$ = 0.02 mbar sample the o-HZO(111) peak is weak. The intensity of this peak increases with deposition pressure (Figure 3b). The m-HZO(002) reflection, barely visible in Figure 3a, can be observed in 2θ-χ frames (Figures 3c and 3d). The intensity of the elongated m-HZO(002) spot is higher in the $PO_2$ = 0.02 mbar film than in the $PO_2$ = 0.2 mbar one. Thus, lowering pressure increases the monoclinic phase and reduces the orthorhombic phase. Oxygen pressure has also an important effect on the lattice strain of the orthorhombic phase. The o-HZO(111) peak (Figure 3a) shifts towards lower angles by reducing deposition pressure. The dependence of $d_{o-HZO(111)}$ with $PO_2$ (Figure 3e) shows that increasing deposition pressure from 0.02 mbar to 0.2 mbar the interplanar spacing decreases from 2.986 Å to 2.954 Å (1.07% contraction).

Figures 2d-f show the influence of the deposition pressure on surface morphology. The 0.01 mbar film (Figure 2d) presents terraces around 100 nm wide, and similar terraces and steps morphology is observed in most of the samples in this series (Figure S4-2). This is not the case of the film deposited at the highest pressure of 0.2 mbar (Figure 3e), where high density of islands increases the roughness to about 0.6 nm. The dependence of the rms roughness on pressure (Figure 3f) reflects the surface roughening with deposition pressure.

The XRD θ-2θ scans of films of varying thickness (t-series) are presented in Figure 4a. The o-HZO(111) peak becomes narrower and more intense when increasing thickness (Figure 4b). The inset shows the linear scaling of the width of this XRD reflection with the reciprocal of the thickness. It signals, according to the Scherrer equation,[30] that epitaxial o-HZO (111) crystals grow across the entire film thickness. The m-HZO(002) peak is seen in films thicker than 10 nm, and the 2θ-χ frames corresponding to the t = 4.6 nm (Figure 4c) and 36.6 nm (Figure 4d) films evidence an increasing fraction of the monoclinic phase respect the orthorhombic with thickness. Whereas the monoclinic phase is not detected in the t = 4.6 nm (Figure 4c), the thickest film (Figure 4d) shows a high intensity m-HZO(002) spot elongated along χ and a weaker m-HZO(-111) spot at 2θ around 28.5°. The 2θ-χ frames of the t = 4.6 nm and thicker films are shown in Figure S5. The dependence on thickness (Figure S5) of the summed intensity area of m-HZO(-111) and m-HZO(002) reflections, normalized to the intensity area of o-HZO(111), shows the progressive increase of monoclinic phase respect to the orthorhombic one with thickness. On the other hand, the orthorhombic phase shows important reduction of the out-of-plane lattice parameter with thickness (Figure 4e), decreasing the interplanar spacing $d_{o-HZO(111)}$ from 3.035 Å to 2.964 Å (2.3% contraction) as thickness increases from 2.3 to 9.2 nm, and presenting little variation in thicker films.

The dependence of the surface morphology on thickness is summarized in Figures 2g-i, and topographic images of all films in the t-series are in Figure S4-3. The morphology of the thinnest film, t = 2.3 nm, shows terraces and steps (Figure 2g), with low rms surface roughness of 0.26 nm. Roughness increases with thickness in films thicker than 10 nm (Figure 3i), up to rms = 0.8 nm in the t = 36.6 nm film. In spite of the higher roughness of this film, morphology of terraces and steps is observed (Figure 3h).

Ferroelectric polarization loops for samples of $T_s$-series and $PO_2$-series are presented in Figure 5a and 5b, respectively. There is hysteresis in all the samples, and there are not wake-up effects as often observed in polycrystalline films.[3,8,28,31,32] The dependence of the ferroelectric properties on deposition conditions can be inferred from Figures 5c and 5d, where the remnant polarization ($P_r$) and the coercive voltage ($V_C$) are plotted as a function of $T_s$ and $PO_2$, respectively. $P_r$ increases with $T_s$ up to around 20 $\mu C/cm^2$ at $T_s$ = 825 °C, and it increases with $PO_2$, strongly for low pressures, from very low polarization up to around 20 $\mu C/cm^2$ for deposition pressure around 0.1 mbar. $V_C$ shows similar trends to $P_r$ with values always below around 3 V. In both series of samples, imprint electric field is present, which produces a shift towards the positives voltage, always smaller than 0.4 V (around 400 kV/cm). Leakage current at several electric fields for all the samples of $T_s$- and $PO_2$-series is shown in Figures 5e and 5f, respectively (leakage curves are presented in Figure S6). The leakage current decreases more than one order of magnitude with $T_s$, and it increases more than three orders of magnitudes with $PO_2$. The leakage of the $PO_2$ = 0.02 mbar film is around $2 \times 10^{-7}$ A/cm² at 1 MV/cm (whereas the 0.01 mbar sample was too insulating for a reliable measurement). The dependence shown in Figure 5f suggests that leakage in this range of $PO_2$ is not dominated by oxygen vacancies. Boundaries between monoclinic and orthorhombic grains and/or crystal domains can present high electrical conductivity. The orthorhombic phase increases with $PO_2$, and an eventual increase in boundaries density could cause larger leakage. Beyond the leakage mechanisms, from the experimental dependences of both polarization and leakage on $T_s$ and $PO_2$, it is concluded that high $T_s$ is convenient for high polarization and low leakage, whereas $PO_2$ in the 0.05-0.1 mbar range is optimal for good combination of large polarization and low leakage.

Ferroelectric P-E hysteresis loops for the samples of t-series are presented in Figure 6a. Films thicker than 4 nm show ferroelectric hysteresis. In the thinnest film (t=2.3 nm), reliable polarization value was not extracted due to the high leakage current contribution. The dependence of remnant polarization on thickness (Figure 6b) shows that the t = 6.8 nm film has the largest $P_r$, and decreasing the polarization with increasing thickness. Similar peaky dependence of $P_r$ with thickness is usual in polycrystalline hafnia films.[3,5,27,33] The thickness dependence of remnant polarization reported for polycrystalline $Hf_{0.5}Zr_{0.5}O_2$ films is compared in Figure S7 to the dependence of our epitaxial films. Remarkably, the thickness for the largest $P_r$ is shifted from above 10 nm in polycrystalline HZO films to around 7 nm in epitaxial films. Leakage current at several electric fields is plotted as a function of thickness in Figure 6c (the corresponding leakage curves are shown in Figure S6). It is seen that the leakage increases by around two orders of magnitude with reducing thickness, presenting the thicker film remarkably low leakage of around $1\times10^{-7}$ A/cm$^2$ at 1 MV/cm. The coercive voltage $V_C$ increases with thickness (Figure 6d, right-axis). Similar $V_C$ values and thickness dependence are obtained if the dielectric contribution is compensated by subtraction of the slope at high field (Figure S8). On the other hand, the measurement of saturated loops is challenging due to the huge coercive electric field of ferroelectric hafnia[32] and coercive field typically depends on the maximum electric field.[14] The polarization loops in Figure 6a were measured with electric field amplitudes as high as possible, close to the breakdown fields as detailed in Figure S8. The electric field is plotted as a function of the thickness in Figure 6d (left-axis, log scale). The slope of linear fit (red dashed line) to log($E_C$) versus thickness is -0.61, which is in agreement with the scaling value of -2/3.[34] This scaling is often observed in ferroelectric perovskite films.[35,36] It requires good screening of polarization charges by the electrodes, particularly for very thin ferroelectric films.[37] However, this scaling behavior has been not observed in polycrystalline ferroelectric hafnia[5] or even in epitaxial hafnia obtained by annealing of room temperature deposited films.[21] Depolarizing effects due to imperfect screening,[8] dispersion of ferroelectric domains in a dielectric matrix[5] or effects of small domain size even in thick films[21] have been proposed as responsible for the up to now elusive observation of $E_C - t^{-2/3}$ scaling in hafnia. Therefore, the $E_C - t^{-2/3}$ scaling in our films, deposited epitaxially at high temperature, signals the importance of the electrodes and film microstructure, and thus high quality samples are required for accurate control of ferroelectricity.

We have presented the growth window of epitaxial HZO films, which permits tailoring structural and ferroelectric properties of the films. In order to elucidate if there is direct effect of structure (relative orthorhombic phase amount and strain), the remnant polarization has been plotted as a function of the normalized intensity of the o-HZO(111) reflection (Figure 7a) and as a function of the $d_{o-HZO(111)}$ interplanar spacing (Figure 7b). Data corresponding to the $T_s$, $PO_2$ and t series are displayed by black squares, red circles and blue triangles, respectively. The polarization scales with the amount of the relative orthorhombic phase excluding films thicker than 10 nm (Figure 7a). Similar correlation between the orthorhombic phase content and ferroelectric polarization was observed for polycrystalline hafnia.[38] On the other hand, the polarization appears to increase as lower is the out-of-plane lattice parameter (Figure 7b). The thicker films of the thickness series deviate from this dependence, but the strong influence of film thickness on the amount of paraelectric monoclinic phase can hide strain effects. Thus, our results demonstrate flexible engineering of the ferroelectric properties of epitaxial films deposited on a particular substrate by proper selection of deposition parameters.

4. CONCLUSIONS

The growth window of epitaxial stabilization of Hf$_{0.5}$Zr$_{0.5}$O$_2$ films on LSMO/STO(001) has been determined. The deposition parameters and thickness have great impact on the orthorhombic phase amount, and the lattice strain can be varied within a range wider than 3%.

The ferroelectric polarization increases with the amount of orthorhombic phase and is found to be larger as smaller is the out-of-plane lattice parameter, and thus it can be controlled by deposition parameters. The leakage current is also conditioned by the deposition parameters, being lower for higher temperature and particularly for lower oxygen pressure. Remarkably, the $E_C - t^{-2/3}$ scaling of electric coercive field and thickness is found by the first time for ferroelectric hafnium oxide, even for films thinner than 5 nm. The growth window map is an important tool for further studies on epitaxial films, for example to get unravel the individual contributions of relative amount of the orthorhombic phase and elastic strain effects on ferroelectric properties.

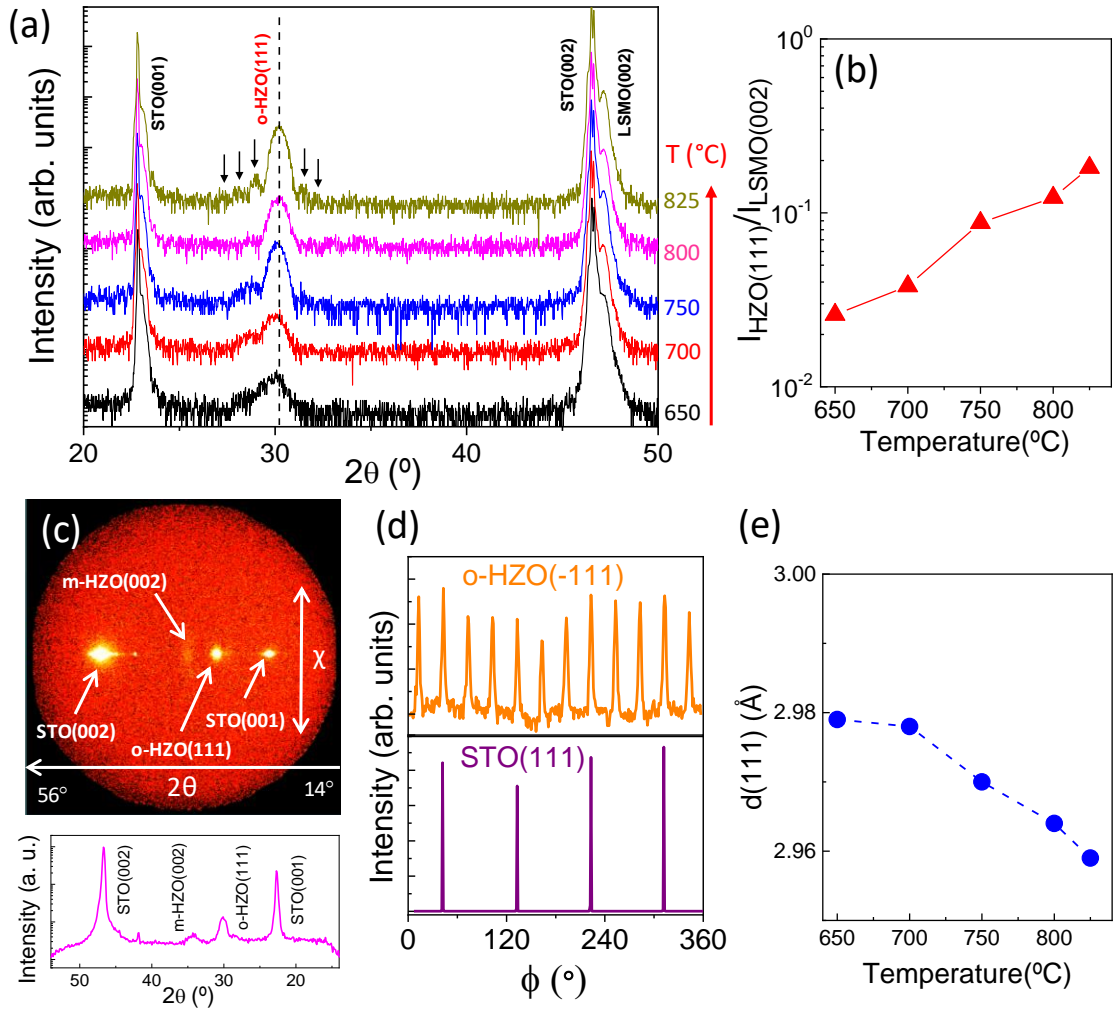

**Figure 1**: (a) XRD θ-2θ scans of HZO films deposited from $T_s$ = 650 °C to 825 °C. The vertical dashed line marks the position of the o-HZO(111) reflection in the $T_s$ = 825 °C film, and vertical arrows mark Laue fringes. (b) Intensity of o-HZO(111) normalized to LSMO(002), plotted as a function of $T_s$. (c) XRD 2θ-χ frame of the $T_s$ = 800 °C film, and θ-2θ scan integrated +/- 5° around χ = 0°. (d) XRD φ-scans around o-HZO(-111) and STO(111) reflections. (e) Dependence on $d_{o\text{-}HZO(111)}$ interplanar spacing with $T_s$.

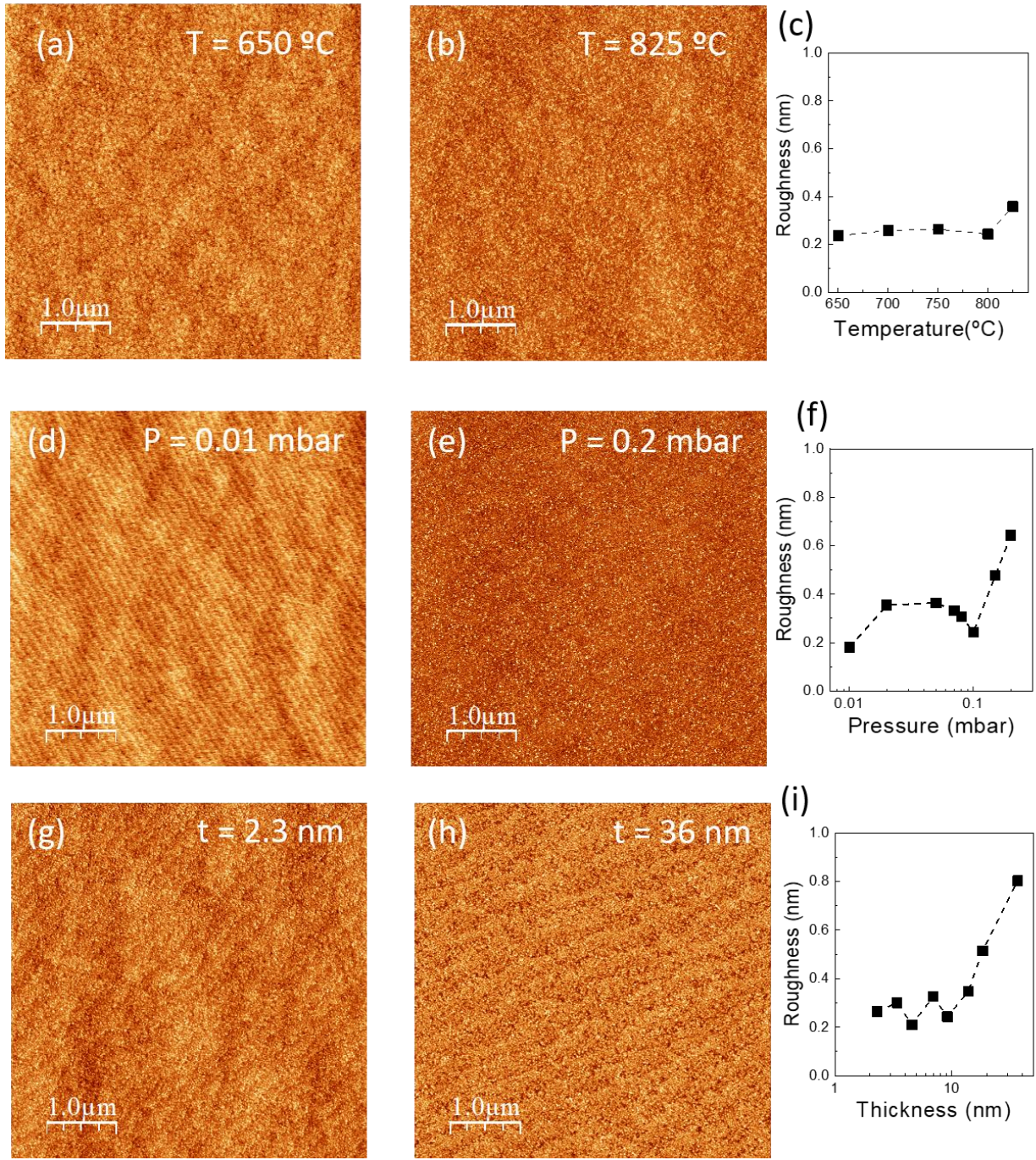

**Figure 2**: AFM topographic 5 µm x 5 µm images of HZO films: deposited at $T_s$ = 650 °C (a) and 825 °C (b); deposited at $PO_2$ = 0.01 mbar (d) and 0.2 mbar (e); and of thickness t = 2.3 nm (g) and 36.6 nm (h). Dependences of root mean square (rms) roughness on $T_s$ (c), $PO_2$ (f) and thickness (i).

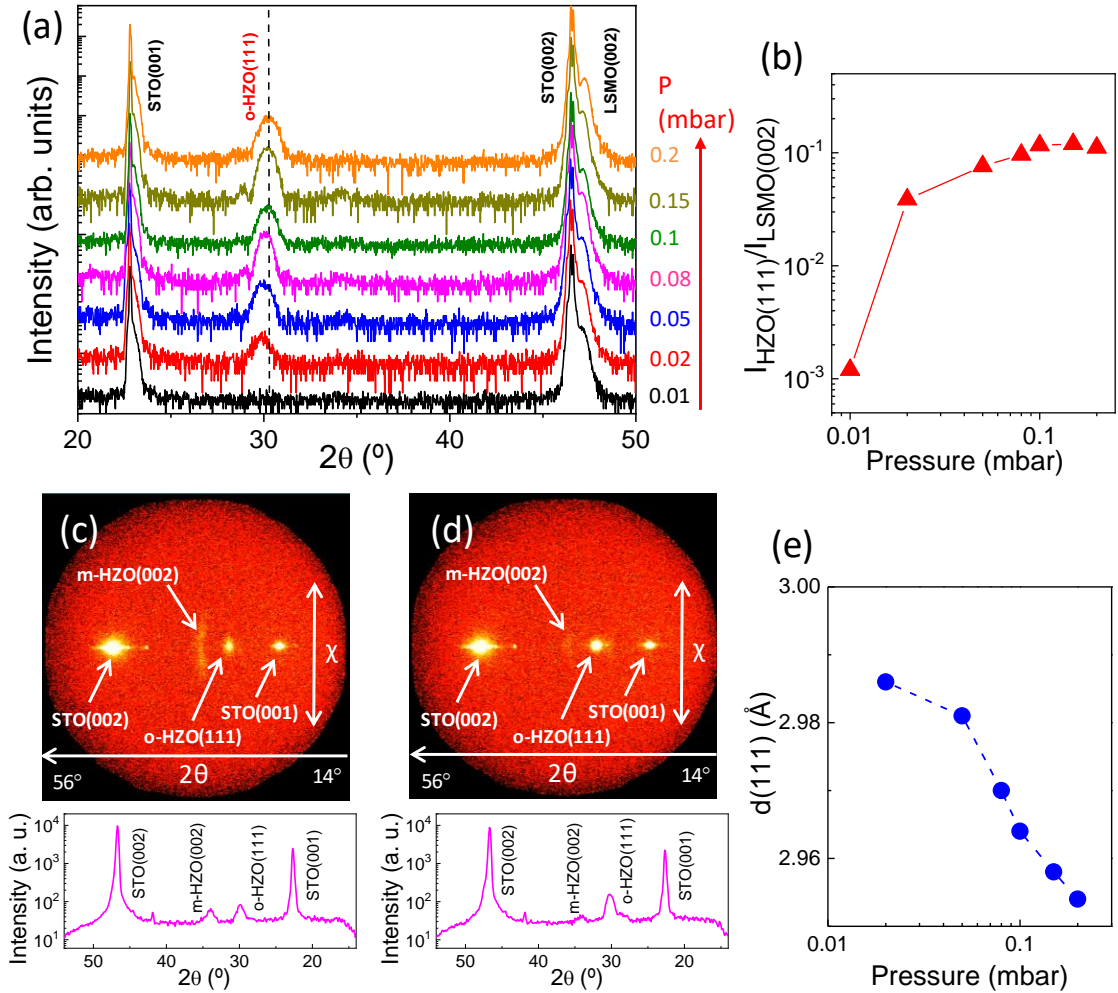

**Figure 3**: (a) XRD θ-2θ scans of HZO films deposited from $PO_2$ = 0.01 mbar to 0.2 mbar. The vertical dashed line marks the position of the o-HZO(111) reflection in the $PO_2$ = 0.2 mbar film. (b) Intensity of o-HZO(111) normalized to LSMO(002), plotted as a function of $PO_2$. XRD 2θ-χ frame of the $PO_2$ = 0.02 mbar (c) and 0.2 mbar (d) films, and corresponding θ-2θ scans integrated +/- 5° around χ = 0 °. (e) Dependence on $d_{o-HZO(111)}$ interplanar spacing with $PO_2$.

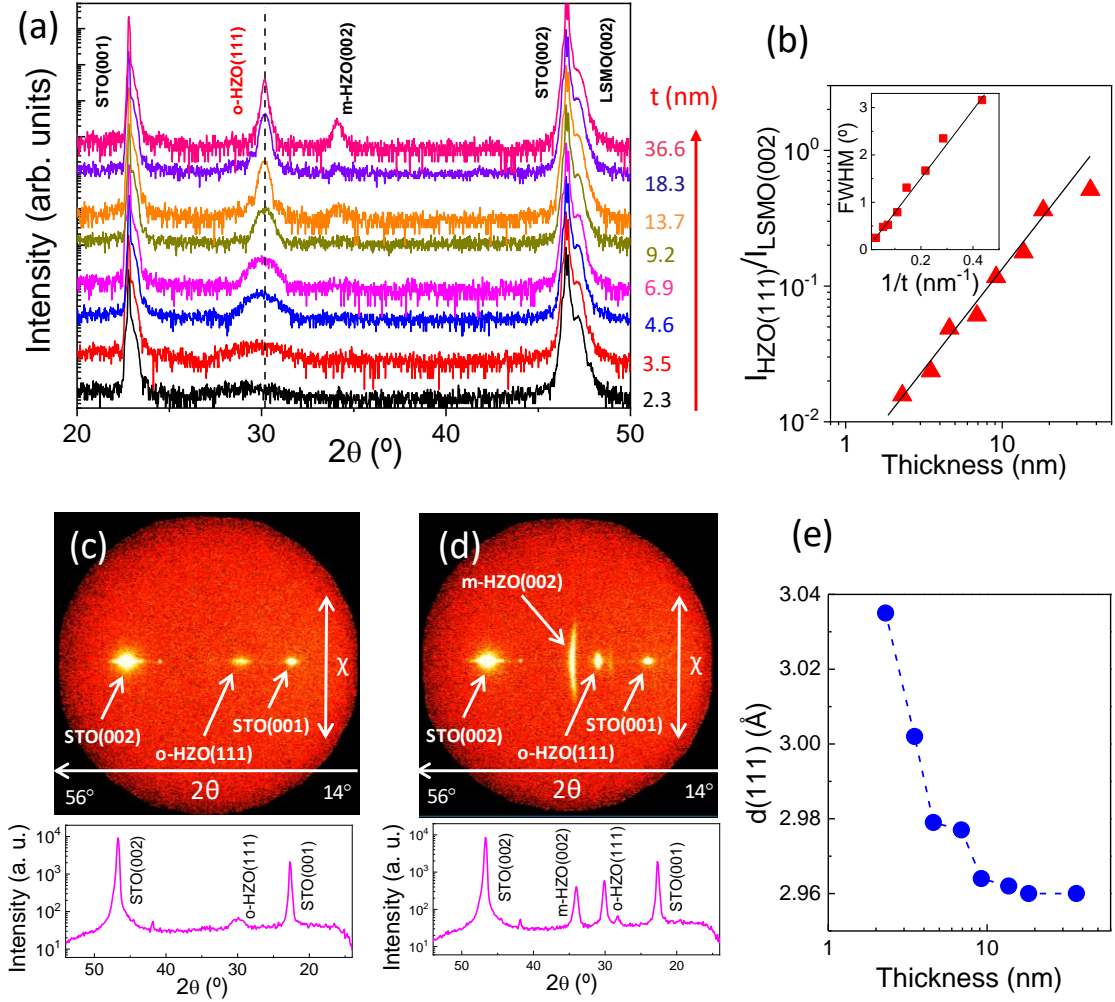

**Figure 4**: (a) XRD θ-2θ scans of HZO films of varying thickness from t = 2.3 nm to 36.6 nm. The vertical dashed line marks the position of the o-HZO(111) reflection in the t = 36.6 nm film. (b) Intensity of o-HZO(111) normalized to LSMO(002), plotted as a function of thickness. Inset: full-width at half-maximum (FWHM) of the o-HZO(111) peak as a function of the reciprocal of film thickness. XRD 2θ-χ frame of the t = 4.6 nm (c) and t = 36.6 nm (d) films, and corresponding θ-2θ scans integrated +/- 5° around χ = 0 °. (e) Dependence on $d_{o-HZO(111)}$ interplanar spacing with film thickness.

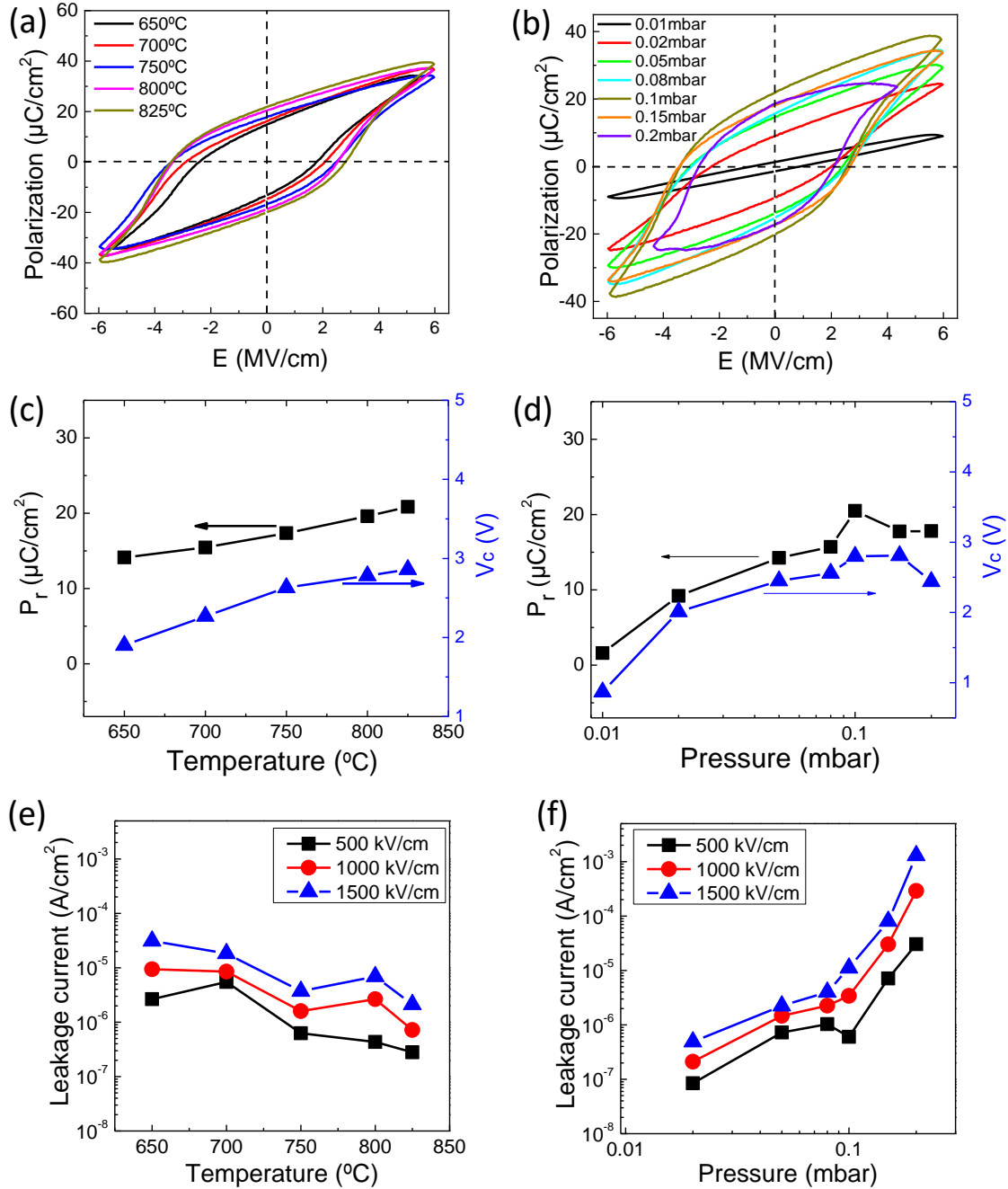

**Figure 5:** (a, b) Polarization – electric field loops for the $T_s$-series and $PO_2$-series, respectively. (c, d) Dependence of $P_r$ and $V_C$ on $T_s$ and $PO_2$ for the $T_s$-series and $PO_2$-series, respectively. Leakage current at the indicated electric fields as a function of $T_s$ (e) and $PO_2$ (f).

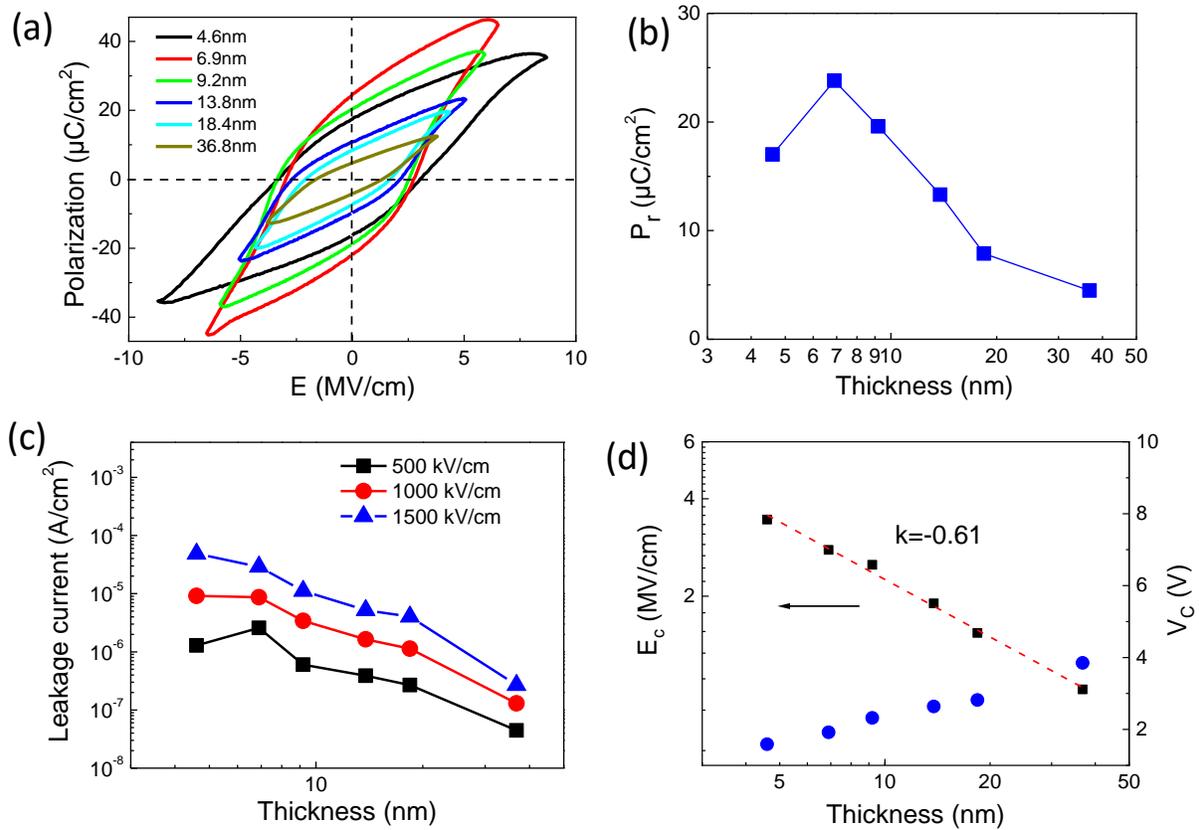

**Figure 6:** (a) Polarization – electric field loops for the samples of the thickness series. (b) $P_r$ dependence on thickness. (c) Leakage current at the indicated electric fields as a function of thickness. (d) $E_C$ (black squares) and $V_C$ (blue circles) dependences on thickness. The red dashed line is a linear fit with slope -0.61, compatible with $E_C - t^{-2/3}$ scaling.

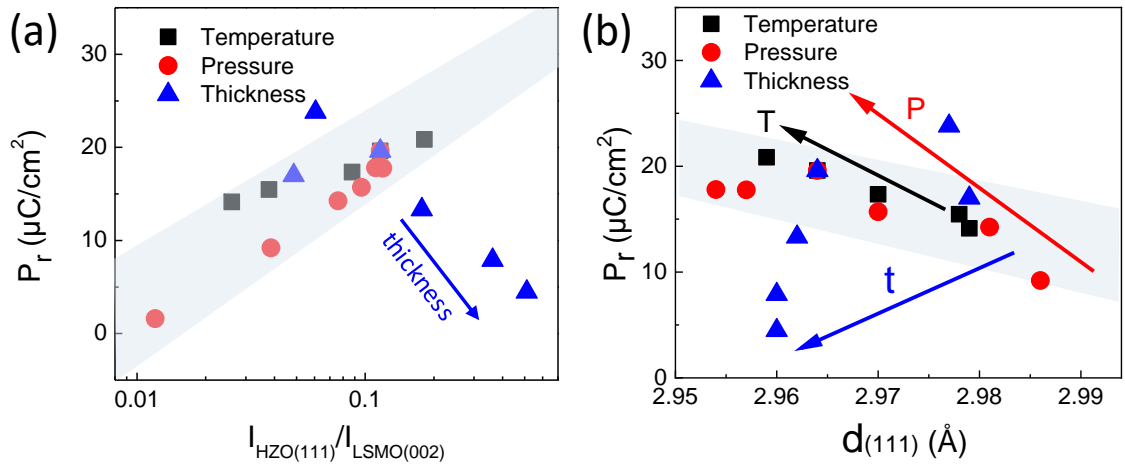

**Figure 7:** $P_r$ plotted against the intensity of the HZO(111) peak normalized to LSMO(002) one (a) and against the out-of-plane lattice parameter of o–HZO, d(111) (b). Black squares, red circles and blue triangles correspond to samples of $T_s$ series, $PO_2$ series, and thickness series, respectively.

ASSOCIATED CONTENT

**Supporting Information**. Schematic sowing the three series of films. Thickness of films deposited under different oxygen pressure. Simulation of Laue interference peaks. Surface morphology of all films. XRD 2D frames of films of varying thickness. Leakage curves. Dependence of remnant polarization with film thickness: comparison with polycrystalline films. Compensation of the dielectric contribution. Polarization loops measured varying the maximum field.

AUTHOR INFORMATION

**Corresponding Author**

*Email: fsanchez@icmab.es

ACKNOWLEDGMENTS

Financial support from the Spanish Ministry of Economy, Competitiveness and Universities, through the "Severo Ochoa" Programme for Centres of Excellence in R&D (SEV-2015-0496) and the MAT2017-85232-R (AEI/FEDER, EU), MAT2014-56063-C2-1-R, and MAT2015-73839-JIN projects, and from Generalitat de Catalunya (2017 SGR 1377) is acknowledged. IF acknowledges Ramón y Cajal contract RYC-2017-22531. JL is financially supported by China

# Supporting Information

# Growth Window of Ferroelectric Epitaxial $Hf_{0.5}Zr_{0.5}O_2$ Thin Films

*Jike Lyu, Ignasi Fina, Raul Solanas, Josep Fontcuberta, and Florencio Sánchez\**

Institut de Ciència de Materials de Barcelona (ICMAB-CSIC), Campus UAB, Bellaterra 08193, Barcelona, Spain

**Schematic showing the three series of films.**

Three series of $Hr_{0.5}Zr_{0.5}O_2$ (HZO) films were grown by pulsed laser deposition on $La_{2/3}Sr_{1/3}MnO_3/SrTiO_3(001)$. A film, common in the three series, was deposited at $T_s$ = 800 °C and $PO_2$ = 0.1 mbar with 800 laser pulses (thickness 9.2 nm). The three series are:

- $T_s$-series: HZO was deposited varying $T_s$ from 650 to 825 °C.
- $PO_2$-series: HZO was deposited varying $PO_2$ in the 0.01-0.2 mbar range.
- Thickness series: HZO was deposited the number of laser pulses in the 200 – 3600 range.

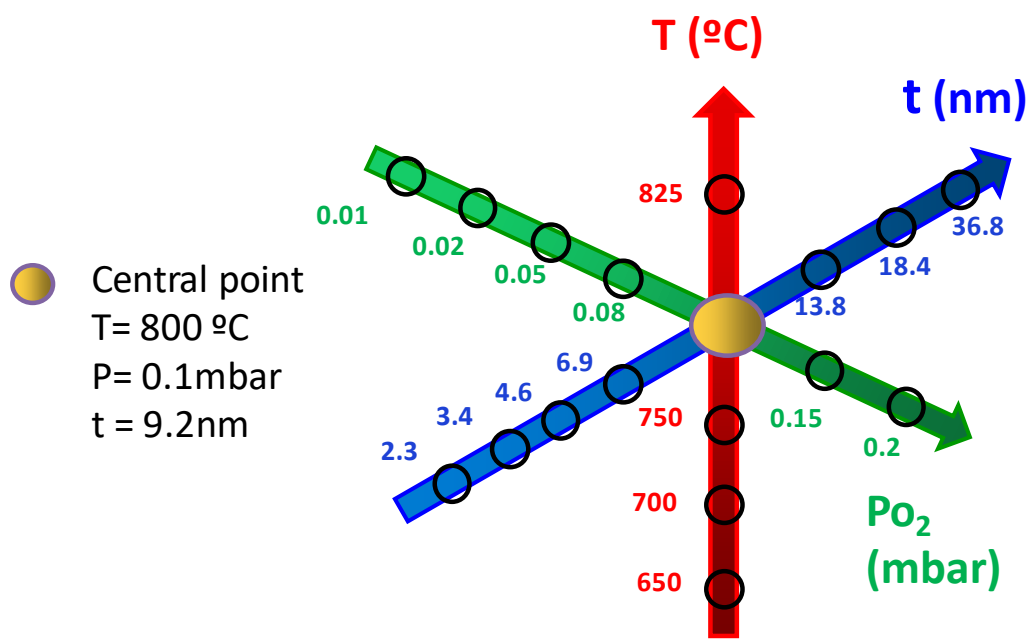

**Figure S1**: Schematic of the three series of HZO.

**Thickness of films deposited under different oxygen pressure.**

The growth rate of HZO was calibrated by X-ray reflectometry of films deposited at 800 °C on bare SrTiO$_3$(001). However, growth rate in pulsed laser deposition can depend on oxygen pressure. Thus, in order to have a direct measurement of thickness of HZO films grown on La$_{2/3}$Sr$_{1/3}$MnO$_3$/SrTiO$_3$(001) at different oxygen pressure, simulations of the Laue interference peaks in the X-ray diffraction (XRD) patterns were done (Figure S2a). The estimated thickness (open blue circles) and growth rate (solid black squares) is shown in Figure S2-b. It is found little dependence on oxygen pressure in the 0.02 – 0.2 mbar range. Slightly larger growth rate of the film grown at 0.08 mbar is due to energy per pulse higher than in the other films in the series.

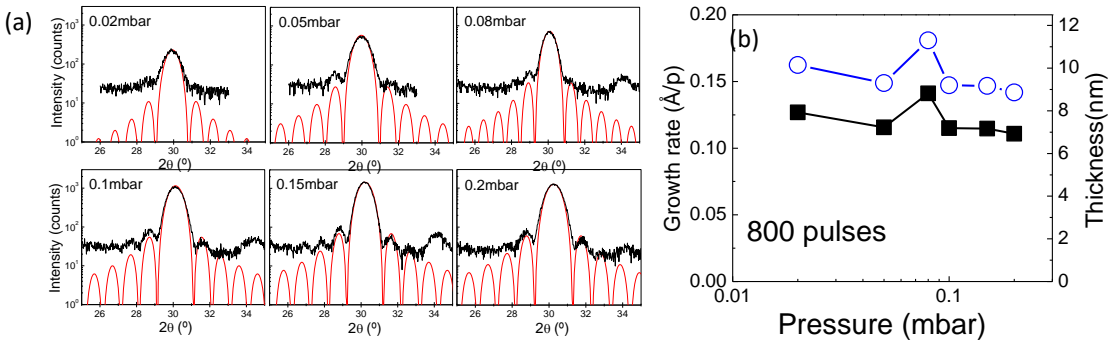

**Figure S2**: (a) XRD patterns of films deposited at different pressure (pressure indicated in each pattern). (b) HZO thickness (left axis, open blue circles) and growth rate (right axis, solid black squares).

**Simulation of Laue interference peaks.**

XRD pattern of the film deposited at T$_s$ = 800 °C and PO$_2$ = 0.1 mbar with 800 (a) and 1600 (b) laser pulses. Measurements were conducted using different diffractometers, with better signal-noise ratio in (b). The red line are simulations of the Laue interference fringes. In the 800 laser pulses film, the fit was done being the o-HZO(111) peak at 2θ = 30.12° and the HZO thickness 92 Å. The respective values for the 1600 laser pulses film are 2θ = 30.152° and 184 Å.

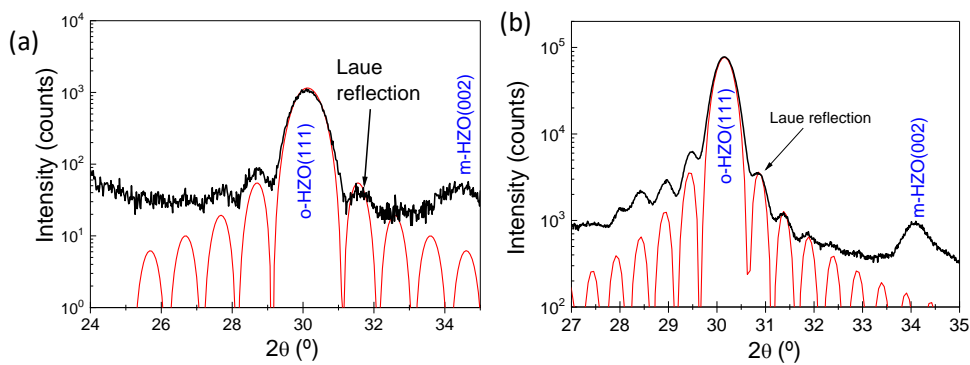

**Figure S3**: XRD pattern of the film deposited at $T_s = 800\ °C$ and $PO_2 = 0.1$ mbar with 800 (a) and 1600 (b) laser pulses. Red lines are simulations of the Laue interference fringes.

**Surface morphology**

The surface morphology of all samples was characterized by topographic atomic force microscopy (AFM). Topographic 5 µm x 5 µm images with height profiles of the films in the $T_s$, $PO_2$, and thickness series are presented in Figures S4-1, S4-2, and S4-3, respectively. The rms roughness is indicated in each image.

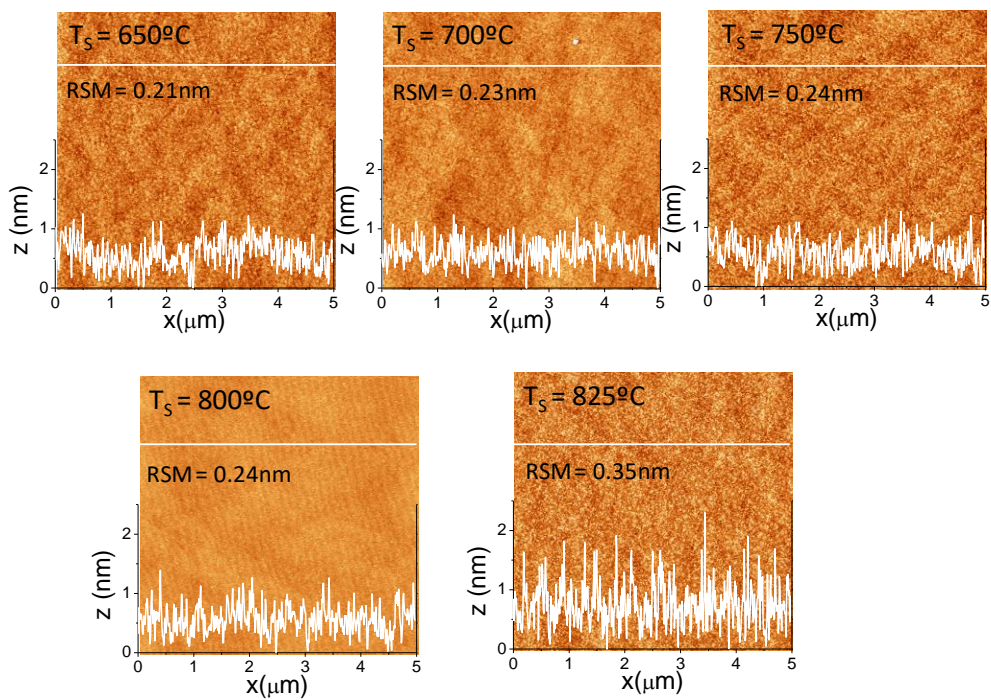

**Figure S4-1**: Topographic AFM images of the HZO films in the $T_s$ series. A height profile along the horizontal marked line is shown in the bottom of each image. The rms roughness of each image is indicated.

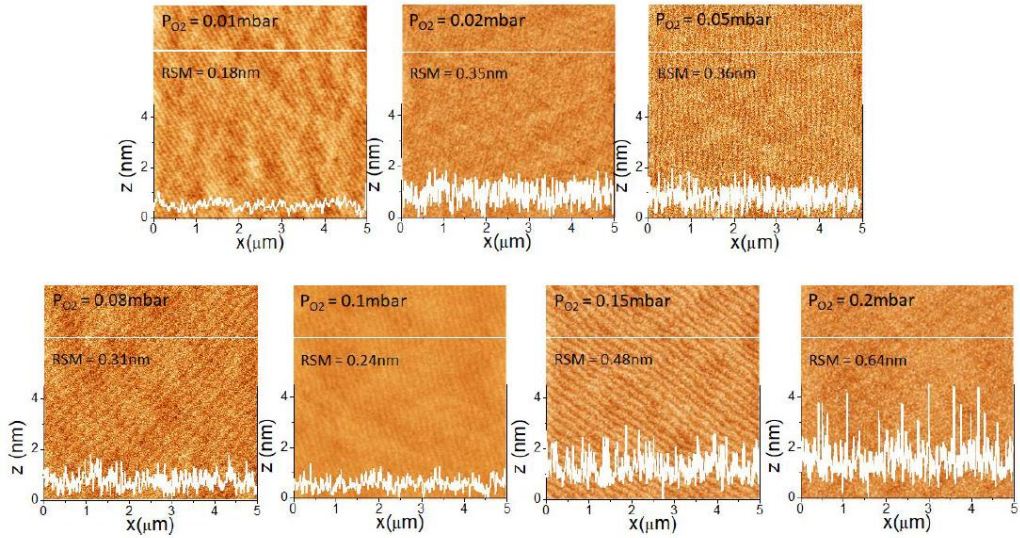

**Figure S4-2**: Topographic AFM images of the HZO films in the $P_{O_2}$ series. A height profile along the horizontal marked line is shown in the bottom of each image. The rms roughness of each image is indicated.

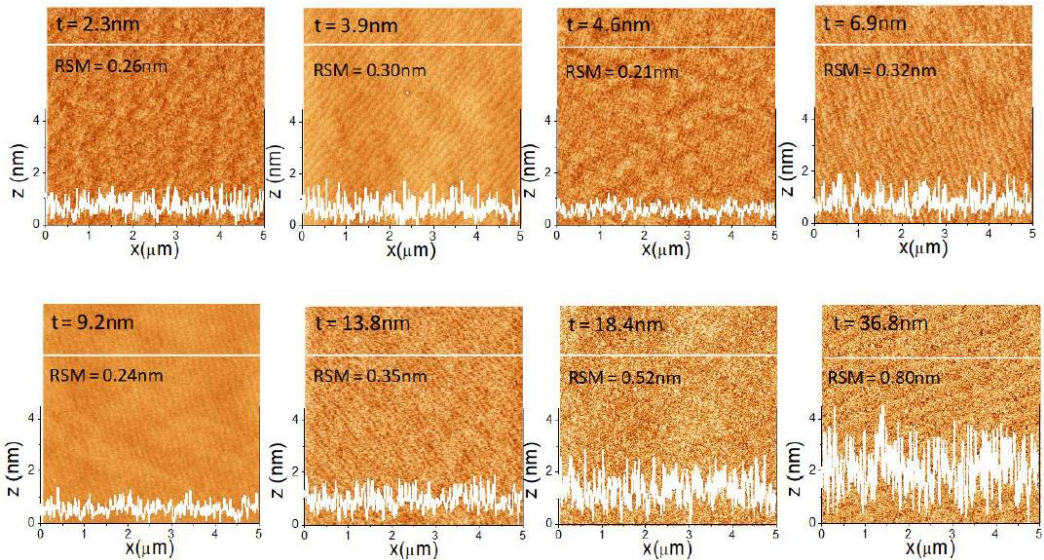

**Figure S4-3**: Topographic AFM images of the HZO films in the thickness series. A height profile along the horizontal marked line is shown in the bottom of each image. The rms roughness of each image is indicated.

## XRD 2D frames of films of varying thickness

XRD 2θ-χ frames of films of thickness t from 4.6 nm to 36.6 nm are presented in Figure S5. The frames show very high increase of intensity of monoclinic (-111) and (002) reflections with thickness. The intensity area of the monoclinic and orthorhombic reflections has been integrated and the fraction between the area of monoclinic and orthorhombic spots is plotted against thickness in the bottom panel.

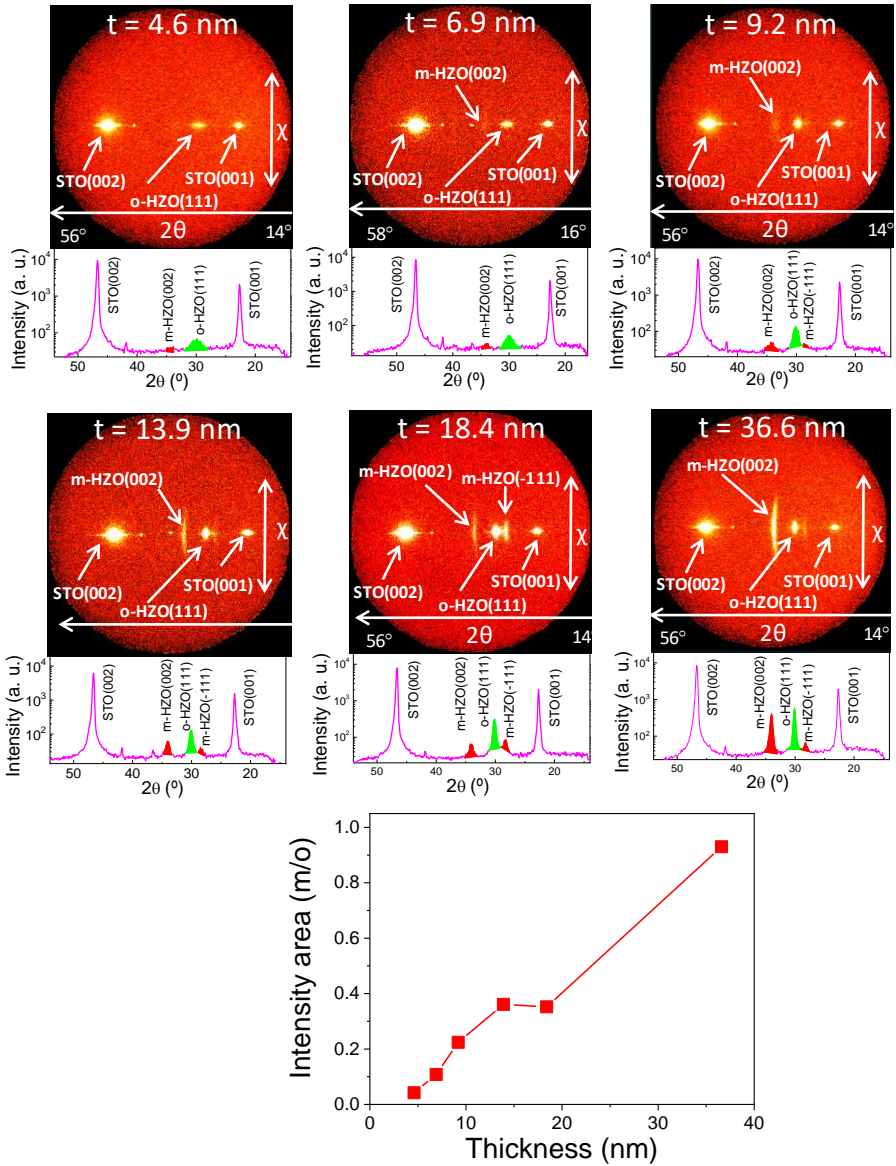

**Figure S5**: XRD 2θ-χ frames of films of varying thickness (indicated in the top of each frame). The 2θ scan below each frame has been obtained by integration in χ from -5 to +5°. The area of m-HZO(1-111) and m-HZO(002) peaks is colored in red, and the area of o-HZO(111) peak in green. Bottom panel: ratio between intensity area of monoclinic HZO (sum of m-HZO(-111) and m-HZO(002) areas) and orthorhombic HZO (o-HZO(111) area) plotted against thickness.

**Leakage curves**

The leakage curves of all HZO films in the $T_s$, $PO_2$ and thickness series are presented. Leakage depends on the substrate temperature and films thickness, and particularly on the oxygen pressure.

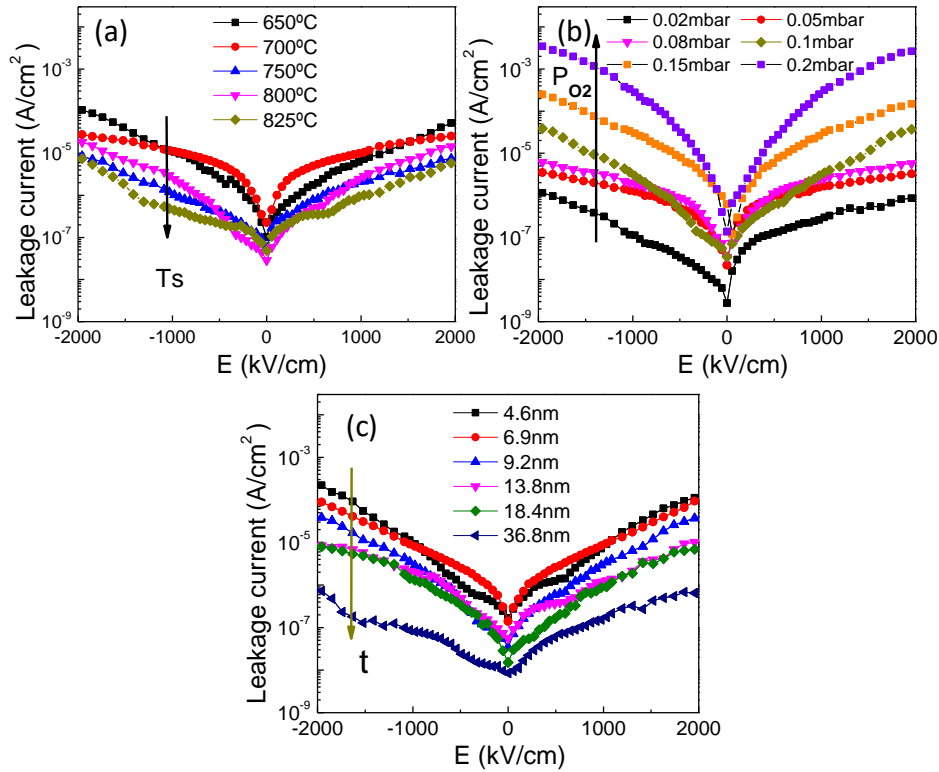

**Figure S6**: Current density – electric field characteristics for the HZO films in the substrate temperature series (a), oxygen pressure series (b), and thickness series (c).

**Dependence of remnant polarization with film thickness: comparison with polycrystalline films**

The thickness dependence of the remnant polarization of the epitaxial HZO films is compared with polycrystalline films (data from literature) having same chemical composition $Hr_{0.5}Zr_{0.5}O_2$ (similar dependences are reported for other dopants). The epitaxial films show a maximum of polarization similarly as the polycrystalline ones, but with a significant shift towards lower thickness (presenting epitaxial films around 7 nm the largest polarization).

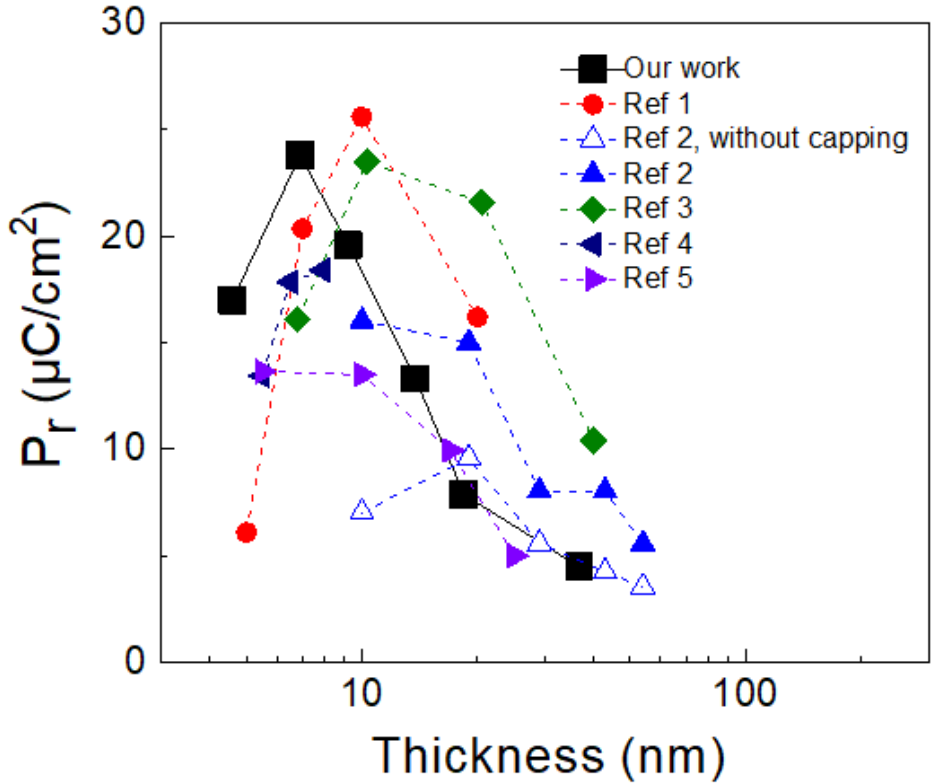

**Figure S7**: Remnant polarization of $Hr_{0.5}Zr_{0.5}O_2$ films as a function of thickness. Black solid squares correspond to the epitaxial films reported here (thickness series). Other symbols (see label in Figure) correspond to polycrystalline $Hr_{0.5}Zr_{0.5}O_2$ films reported in the indicated references.

**Compensation of the dielectric contribution**

The electric susceptibility contribution of the loops can be removed by subtraction of the constant slope at high field. Figure S7a shows the loop of the t = 9.2 nm sample ($T_s$ = 800 °C, 0.1 mbar) before and after dielectric compensation. In the Figure we also show the dependences of coercive electric field and coercive voltage with thickness from loops without (b) and with (d) compensation. It is seen that the dependences are similar, and the slopes (k = -0.61 and k = -0.59) are in both cases compatible with $E_c - t^{-2/3}$ scaling.

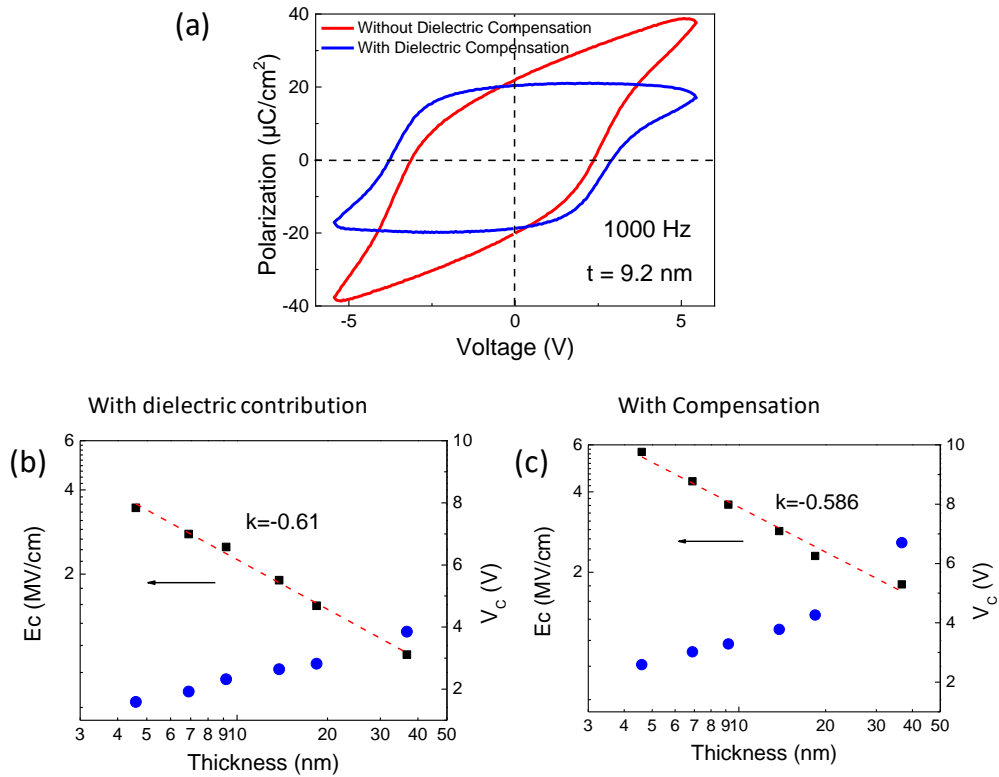

**Figure S8**: (a) Polarization – voltage loops of the t = 9.2 nm sample without (red curve) and after (blue curve) compensation of the dielectric contribution. (b) $E_C$ (black squares) and $V_C$ (blue circles), determined from uncompensated loops, dependences on thickness. (c) Equivalent plot (Figure 6d) determined from compensated loops.

**Polarization loops measured varying the maximum field**

The high coercive fields in ferroelectric $HfO_2$, particularly in epitaxial films, limit the range of electric field that can be applied to measure polarization loops. The polarization loops presented in the paper were measured using electric field amplitudes as high as possible in order to obtain saturated loops. Figure S9 shows loops of the t = 6.9 nm (a) and t = 36.6 nm (b) films measured at varying applied field. Breakdown field decreases with thickness, being around 6.8 and 4.1 MV/cm for the t = 6.9 nm (c) and t = 36.6 nm (d) samples. Figures S9e and S9f show the dependence of coercive field with maximum applied field for the t = 6.9 nm (e) and t = 36.6 nm (f) samples.

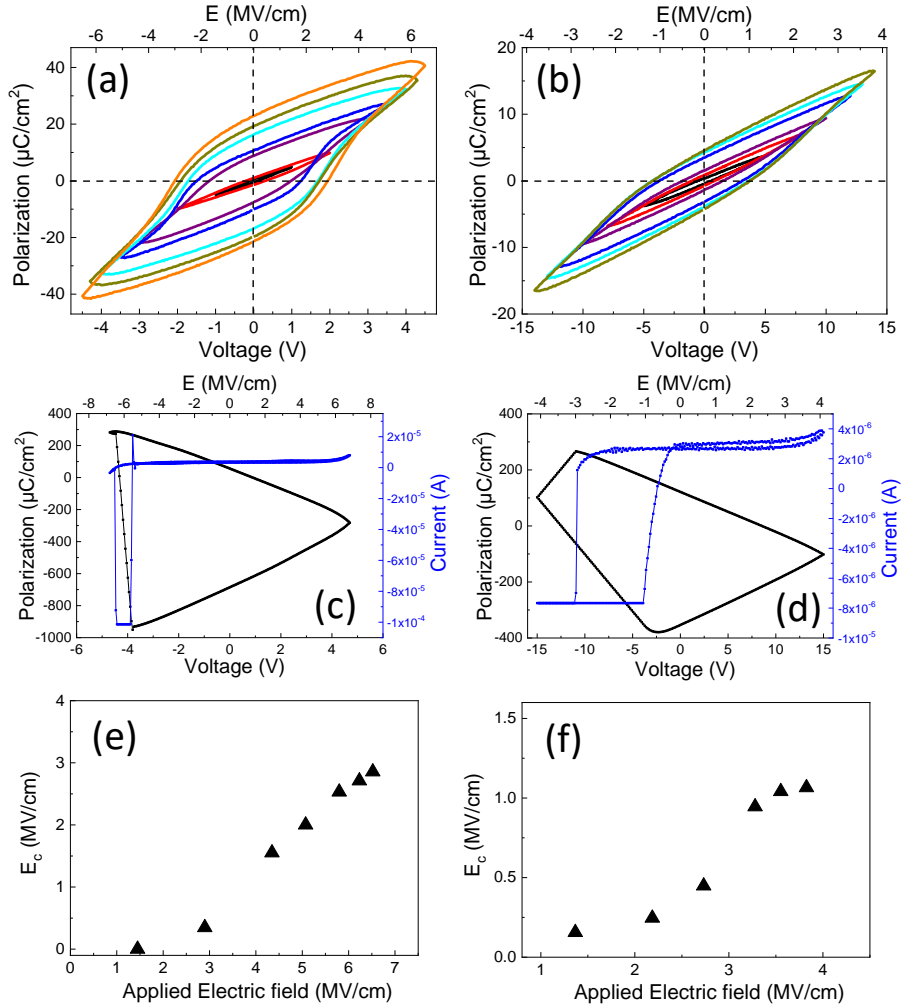

**Figure S9**: Polarization loops of the t = 6.9 nm (a) and t = 36.6 nm (b) films. Breakdown of a capacitor in the t = 6.9 nm (c) and t = 36.6 nm (d) samples at applied fields of E = 6.8 MV/cm and E = 4.1 MV/cm, respectively. Coercive field $E_c$ as a function of the amplitude of the applied field for the t = 6.9 nm (e) and t = 36.6 nm (f) films.